\documentclass[%
 reprint,
superscriptaddress,
 amsmath,amssymb,
 aps,longbibliography,
prb,
]{revtex4-2}

\usepackage{graphicx}
\usepackage{dcolumn}
\usepackage{bm}

\usepackage[bookmarksnumbered,pdfpagelabels=true,plainpages=false,colorlinks=true,linkcolor=blue,citecolor=red,urlcolor=blue]{hyperref}
\usepackage[normalem]{ulem}

\begin{document}

\title{Nonreciprocal Spin-Wave Propagation in Anisotropy-Graded Iron Films Prepared by Nitrogen Implantation}

\author{L. Christienne}
\affiliation{Sorbonne Université, CNRS, Institut des Nanosciences de Paris, INSP, F-75005 Paris, France} 
\author{J. Jiménez-Bustamante}
\affiliation{Departamento de F\'{i}sica, Universidad T\'{e}cnica Federico Santa Mar\'{i}a, Avenida Espa\~{n}a 1680, Valpara\'{i}so, Chile} 
\author{P. Rovillain}
\email{pauline.rovillain@insp.upmc.fr}
\affiliation{Sorbonne Université, CNRS, Institut des Nanosciences de Paris, INSP, F-75005 Paris, France} 
\author{Mahmoud Eddrief}
\affiliation{Sorbonne Université, CNRS, Institut des Nanosciences de Paris, INSP, F-75005 Paris, France} \author{Yunlin Zheng}
\affiliation{Sorbonne Université, CNRS, Institut des Nanosciences de Paris, INSP, F-75005 Paris, France} 
\author{Franck Fortuna}
\affiliation{Université Paris-Saclay, CNRS, Institut des Sciences Moléculaires d’Orsay, 91405 Orsay, France} 
\author{Massimiliano Marangolo}
\affiliation{Sorbonne Université, CNRS, Institut des Nanosciences de Paris, INSP, F-75005 Paris, France} 
\author{M. Madami}
\affiliation{Dipartimento di Fisica e Geologia, Università di Perugia, I-06123 Perugia, Italy} 
\author{R. A. Gallardo}
\affiliation{Departamento de F\'{i}sica, Universidad T\'{e}cnica Federico Santa Mar\'{i}a, Avenida Espa\~{n}a 1680, Valpara\'{i}so, Chile} 
\author{P. Landeros}
\email{pedro.landeros@usm.cl}
\affiliation{Departamento de F\'{i}sica, Universidad T\'{e}cnica Federico Santa Mar\'{i}a, Avenida Espa\~{n}a 1680, Valpara\'{i}so, Chile} 
\author{S. Tacchi}
\email{tacchi@iom.cnr.it}
\affiliation{Istituto Officina dei Materiali del CNR (CNR-IOM), Sede Secondaria di Perugia, c/o Dipartimento di Fisica e Geologia,
Università di Perugia, I-06123 Perugia, Italy}
\date{\today }
\pacs{}
\keywords{spin waves, nonreciprocity, graded anisotropy, ion implantation, magnonics}

\begin{abstract}
Gradual modification of the magnetic properties in ferromagnetic films has recently been proposed as an effective method to channel and control spin waves for the development of new functionalities in magnonic devices. Here, we investigate graded FeN films prepared by low-dose nitrogen implantation of Fe epitaxial thin films. Combining Brillouin light scattering measurements and a spin-wave theoretical approach, we show that nitrogen implantation induces a graded profile of both the in-plane and the perpendicular anisotropies along the film thickness. This graduation leads to a significant modification of the spin-wave spatial localization and generates a marked frequency asymmetry in the spin-wave dispersion. Moreover, we find that the anisotropy profile, and as a consequence the dispersion relation, can be tuned on changing the implantation dose, opening a way for the potential use of the graded Fe-N films in magnonic applications.
\end{abstract}

\maketitle

\section{Introduction}
\label{sec:introduction}

A significant portion of contemporary magnetism research is focused on manipulating the magnetic anisotropy of thin films and nanostructures. This meticulous control holds significant promise for advances in spintronics and magnonics applications, where tailoring both in-plane and out-of-plane magnetic anisotropy is crucial. One notable application is the enhancement of perpendicular magnetic anisotropy (PMA), essential for downsizing devices for higher-capacity storage \cite{Sbiaa2011}. Moreover, PMA has recently emerged as pivotal in stabilizing chiral textures for magnetic recording and data processing \cite{Fert2017}. In the field of magnonics, the fine-tuning of magnetic anisotropy can enable the manipulation of magnetization dynamics and facilitate controlling spin-wave propagation \cite{Barman2021,Flebus2024}. 
Several techniques have been proposed to control magnetic anisotropies, such as substrate-induced stress \cite{Mathews2023}, ion doping during growth \cite{Soumah2018}, voltage-controlled piezoelectric effect \cite{Finizio2014}, interface effects \cite{Sjstedt2002}, and ion implantation \cite{Obry2013}.
The modification of magnetic properties in low-damping materials can manifest in various ways, including uniform, graded, and lateral adjustments. A prime illustration is the uniform Bismuth doping of Y$_3$Fe$_5$O$_{12}$ (BiYIG) films—a material renowned for its low losses and high PMA \cite{Soumah2018,Das24}.
In this context, graded magnetic materials, where physical properties such as saturation magnetization \cite{Mantese05,Sudakar07,Petrov08,Chen14,Fallarino16,Fallarino18,Gallardo2019,Fallarino21,Gallardo22}, exchange constant \cite{Berger08,Graet15,Fallarino17,Kirby18,Fallarino21b}, Dzyaloshinskii-Moriya strength \cite{Kim19,Yershov20,Park22,Liang22,Zhang22,Gallardo24}, or magnetic anisotropy \cite{Suess06,Zhou09,Dumas11,Kirby10,Jiang14,Gladii16,Fallarino20,Szulc24} gradually change along a specific direction, also represent an exciting class of systems that offer a promising approach for guiding and manipulating magnon propagation.
Therefore, the capability to carefully tune the material properties in a gradual manner and on a nanometric scale is essential to engineer spin-wave dispersion relations for specific magnonic applications.

Graduating a magnetic property through a given direction, indeed, breaks the spatial symmetry and, as a consequence, affects both the allowed spin-wave frequencies and the spatial mode profiles.
In particular, it has been theoretically predicted that a graded variation of the magnetic properties along the film thickness causes heterosymmetric mode profiles and a modification of the conventional quantization condition associated with the perpendicular standing spin waves, inducing a significant frequency nonreciprocity of counterpropagating waves \cite{Gallardo2019,Brevis24}.
This latter behavior is currently of great interest in the research area of magnonics since nonreciprocal spin-wave propagation is crucial for the implementation of unidirectional devices, such as spin-wave circulators and diodes, which serve as basic building blocks for magnonic circuits \cite{Jamali13,Reiskarimian2016}. Nonreciprocal spin-wave propagation has been experimentally demonstrated for ferromagnetic bilayers consisting of different materials in direct contact, which can be regarded as a class of graded films with a sharp profile in saturation magnetization \cite{Mruczkiewicz2017,Grassi2020}.

Recently, ion implantation has emerged as a powerful tool for realizing graded systems. This technique allows the introduction of different kinds of atoms with high precision in dosing, thereby enabling a controlled adjustment of the magnetic parameters such as the saturation magnetization or the magnetic anisotropy \cite{Fassbender08,McCord09,Lee20}. In addition, depth control can be finely regulated by modifying the energy of impinging ions. This precise tuning can be accomplished uniformly across large samples or at the nanoscale level through focused ion beam (FIB) implantation.

In this work, we present a detailed study of the spin wave properties in nitrogen-implanted FeN thin films. 
In previous studies, it was demonstrated that nitrogen implantation induces an increase of the out-of-plane anisotropy in Fe films while preserving the film epitaxy and low damping, showing the potential of this methodology for magnonic applications \cite{Amarouche2017,theseGarnier,Camara2017}. 
Here, we investigate the influence of graded anisotropy on the spin-wave dispersion of FeN thin films by means of Brillouin light scattering measurements and theoretical calculations.  We show that low-dose nitrogen implantation induces a gradual increase of the perpendicular anisotropy and a decrease of the in-plane cubic one in the surface region of a Fe film. These anisotropy variations affect the spatial localization of the spin-wave modes and, as a consequence, significantly modify their dispersion relation. In particular, we observed a marked frequency asymmetry between spin-wave modes propagating in opposite directions, which can be tuned by changing the irradiation dose.

\section{Experiments }
\subsection{Sample Preparation and Structural Characterization}
\begin{figure}[hb]
    \centering
    \includegraphics[width=.25\paperwidth]{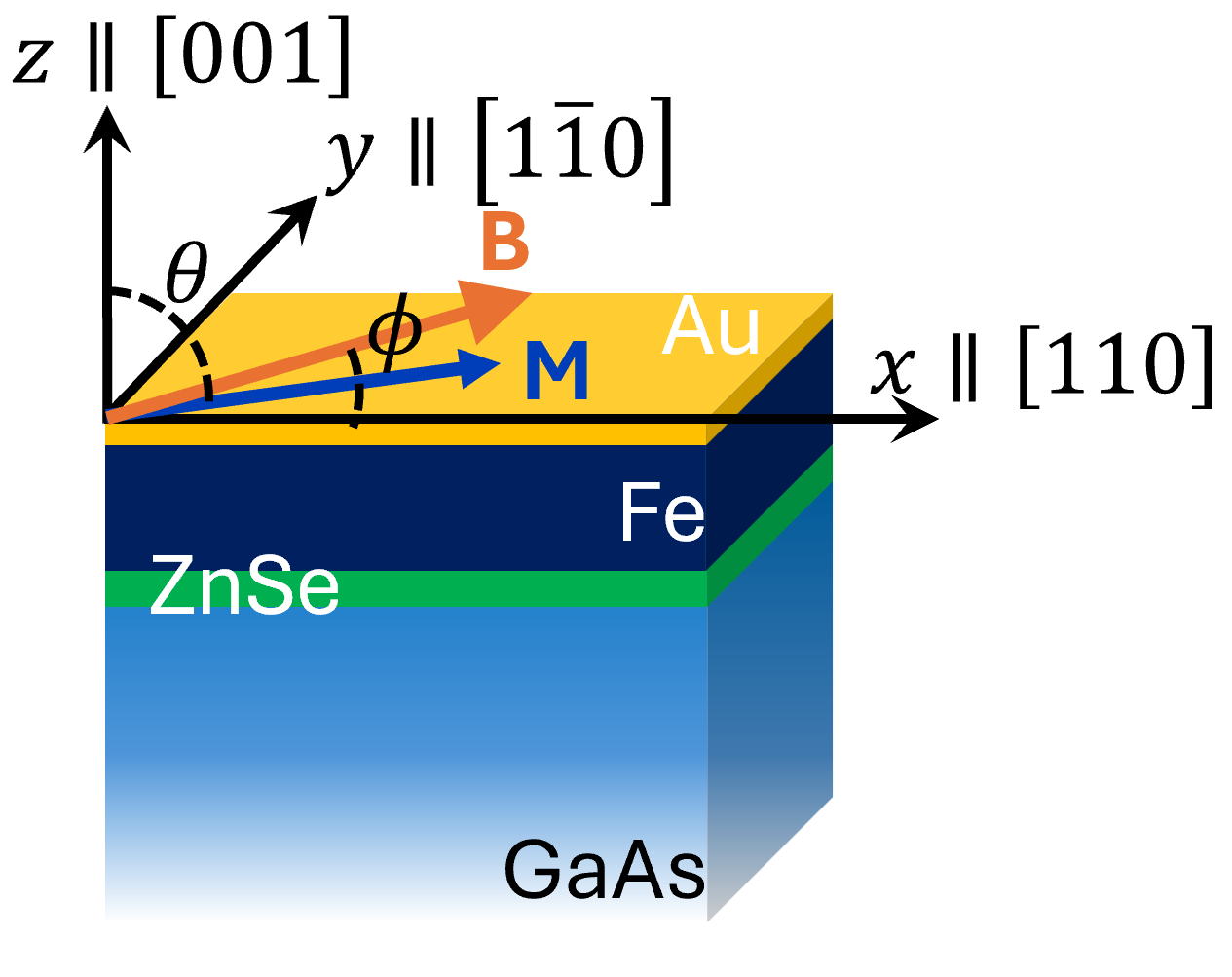}
    \caption{Sketch of the sample: Fe or Fe:N (54 nm
thick) epitaxied on ZnSe(001)/GaAs(001). \textbf{B} is the in-plane applied magnetic field, and \textbf{M} is the magnetization.}
    \label{fig:Coord}
\end{figure}

Iron films having a thickness of 54~nm were grown using molecular beam epitaxy (MBE) on non-doped GaAs (001) wafers (see Fig.~\ref{fig:Coord}). %
\begin{figure}[hb]
    \centering
    \includegraphics[width=.37\paperwidth]{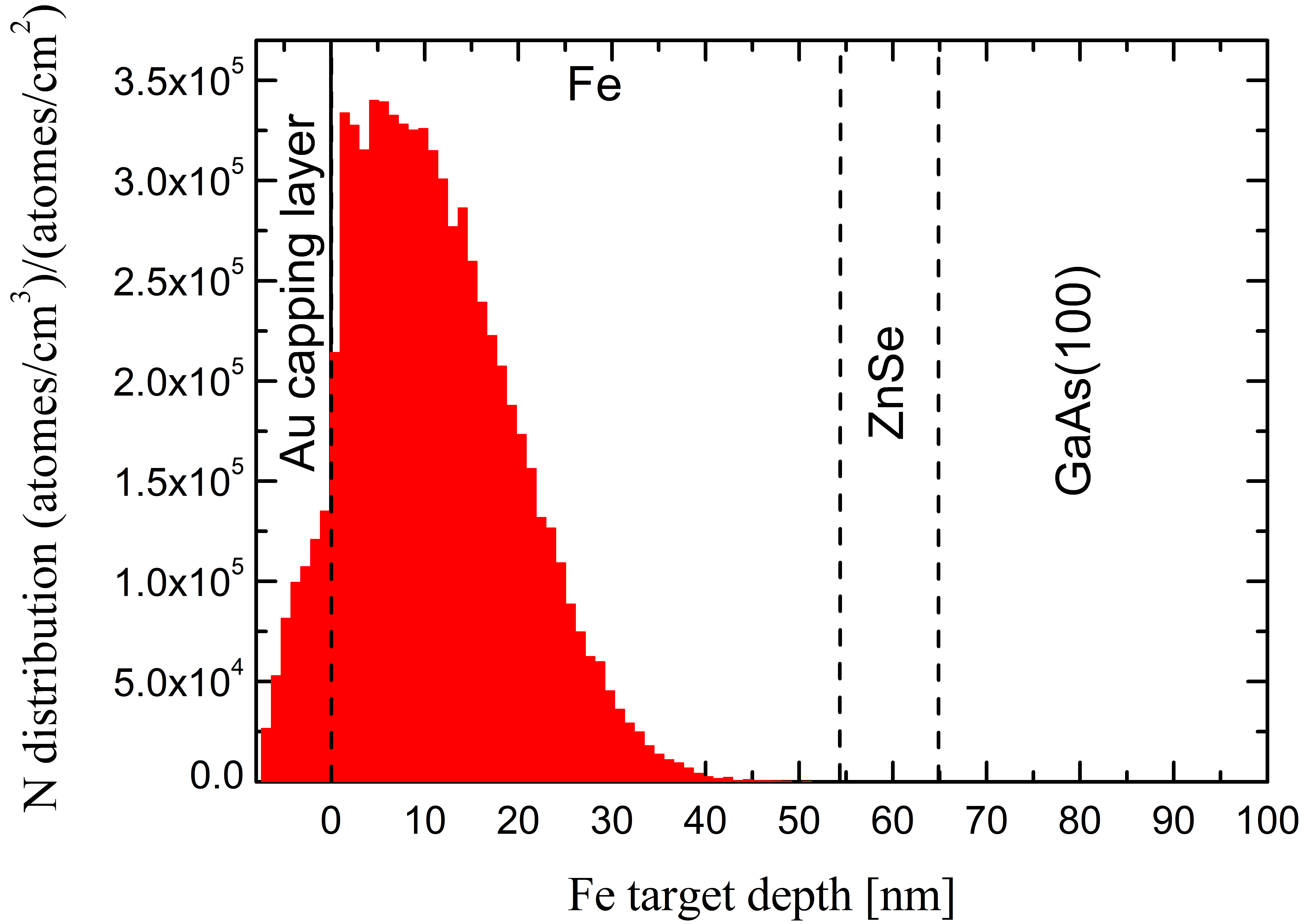}\\
    \caption{TRIM simulation of the implantation with N$_2$ ions accelerated at 26 keV and implanted in a Fe film epitaxied on ZnSe/GaAs(001) substrate. The dose is 1$\times10^{16}$ N$_2$/cm$^2$.} 
    \label{fig:TRIM}
\end{figure}
Prior to Fe deposition, a thin layer (10~nm) of ZnSe was applied to prevent migration \cite{Marangolo2004}, which could create a magnetic ``dead layer". The growth of Fe layers occurred with the in-plane [100]-Fe axis perfectly aligned azimuthally with that of the underlying GaAs substrate, while the [001] axis was out-of-plane. A gold capping layer (8~nm) was deposited in a separate chamber in order to prevent oxidation of the iron layer. 
Samples, measuring about $10\times 5$ mm, were cleaved from the wafers and resized for various measurements. Two of these samples underwent two different implantation doses at the IN2P3 JANNuS platform using a Nier-Bernas 190~kV ion implanter. The acceleration was set at 26~keV, and the chosen doses were $1.0 \text{ and } 1.5 \times10^{16}$ N\textsubscript{2}/cm\textsuperscript{2}. Transport of Ions in Matter (TRIM) simulations provide rough estimates of the implantation depth, as shown in  Fig.~\ref{fig:TRIM}. It is important to note that the N implantation within the Fe film is not homogeneous. N ions are mainly concentrated in the upper part of the Fe overlayer underlying the Au/Fe interface.

\begin{figure}[h]
    \centering
\includegraphics[width=.37\paperwidth]{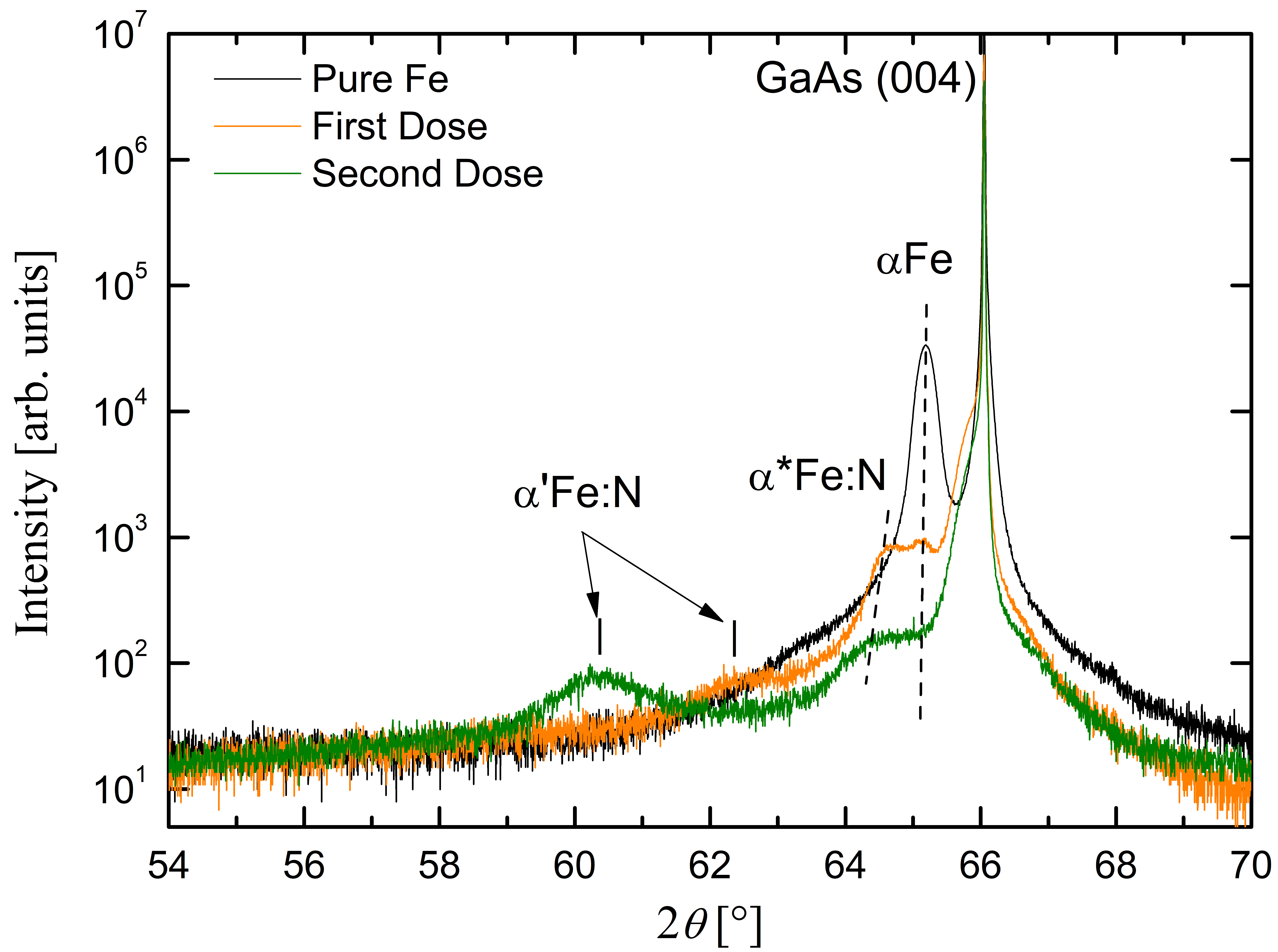}
    \caption{$\theta$-2$\theta$ measurements. Colors correspond to different N\textsubscript{2} doses implanted into Fe films epitaxially grown on ZnSe/GaAs(001); black for pristine Fe, orange for the first dose (1.0 $\times$ 10$^{16}$ N$_2$/cm$^2$) and green for the second dose (1.5 $\times$ 10$^{16}$ N$_2$/cm$^2$). Two distinct Fe:N phases are clearly observed, both exhibiting tetragonal distortion.}
    \label{fig:DRXth2th}
\end{figure}
During the nitrogen implantation, the emergence of Fe$_{1-x}$N$_{x}$ phases with crystalline parameters differing from bcc iron is expected \cite{theseGarnier}. X-ray diffraction measurements are conducted on our samples to determine the present phases and to extract the associated lattice parameters.
Measurements are performed using a Rigaku SmartLab 5-axis diffractometer. From a fit of the X-ray reflectivity oscillations, the mean thickness of the iron layer, $d$, of around 54 nm was obtained. Figure \ref{fig:DRXth2th} shows the $\theta$-$2\theta$ measurements of the (001) plane of the three different samples as pristine Fe, first dose (1.0 $\times$ 10$^{16}$ N$_2$/cm$^2$) and second dose (1.5 $\times$ 10$^{16}$ N$_2$/cm$^2$) irradiated samples. Another sample with a higher dose (2$\times$ 10$^{16}$ N$_2$/cm$^2$) was also studied to put into evidence the tetragonal distortion by a reciprocal space mapping (see Appendix \ref{App.A}). These measurements allow access to the out-of-plane lattice swelling, notably by tracking the peaks identified as $\alpha$\textsuperscript{$\star$}-Fe:N, a bcc-phase with lattice parameters very close to pristine Fe ones. An $\alpha$'-Fe:N phase is also observable, corresponding to a phase trending towards Fe\textsubscript{8}N\textsubscript{1} stoichiometry \cite{Garnier2016,Jack1951,Wriedt1987}. We notice that for $\alpha^\star$-FeN$_x$ and $\alpha$’-Fe$_8$N$_{1-x}$, only diffraction peaks corresponding to crystal planes parallel to the (001) surface of the GaAs substrate are observed. This indicates that the [001] directions of these two phases are parallel to the [001] direction of GaAs. The corresponding in-plane and out-of-plane lattice parameters are reported in Table \ref{tab:parametrec}.
Both N-rich phases exhibit an increase in the lattice parameter \( c \), with a significantly greater expansion observed in the \( \alpha' \)-Fe\(_8\)N\(_{1-x}\) phase, as shown in Table~\ref{tab:parametrec}. Regarding the in-plane parameter, the measured values for both phases are slightly higher than that of bulk Fe. Additionally, we observe that the lattice parameters of the \( \alpha^\star \)-FeN\(_x\) phase are very close to those of Iron, making it effectively Fe-like.

\begin{table}[htb]
\fontsize{10pt}{10pt}
\caption{Lattice parameters.}
\label{tab:parametrec}
\renewcommand{\arraystretch}{2.0} 
\centering
\begin{ruledtabular}
\begin{tabular}{ccccc}
     & \multicolumn{2}{c}{$\alpha$\textsuperscript{$\star$}-Fe:N (Fe-like)}& \multicolumn{2}{c}{$\alpha$'-Fe:N}\\
    \hline
     &  $c$ (\r{A})& $a$ (\r{A})&  $c$ (\r{A})& $a$ (\r{A})\\
    \hline
    Pristine Fe&2.86&2.870 & - & - \\
    First dose & 2.87 & 2.877 & 2.96 & - \\
    Second dose & 2.88 & 2.872 & 3.07 & 2.862 \\
\end{tabular}
\end{ruledtabular}
\end{table}

To illustrate the tetragonal distortion and demonstrate the preservation of pristine Fe epitaxy conditions, the (112) diffraction spot of a sample at higher dose (2.0 $\times$ 10$^{16}$ N$_2$/cm$^2$) along the $Q_x$ ($\parallel$[110]) and $Q_z$ ($\parallel$[001]) directions is mapped. A reciprocal lattice map is presented and discussed in Appendix \ref{App.A}. The results indicate that the epitaxial conditions of the pristine Fe/GaAs(001) remain intact after N$_2$ implantation.  These findings are consistent with those previously reported in Ref.~\onlinecite{theseGarnier} for higher implantation dose.

\subsection{Magnetic Characterization}

\subsubsection{Vibrating Sample Magnetometry}

Vibrating-sample magnetometry (VSM) was employed to measure the magnetization loops of the pristine Fe and implanted samples (see Fig.~\ref{fig:OOPmaille}). A magnetic field up to 3 T was applied perpendicular to the film surface. For each sample, the saturation magnetization $M_{\rm s}$ was first extracted from the measured saturation value. 
Then the out-of-plane anisotropy constant was determined using the Stoner-Wohlfarth method, where the energy density can be written $\epsilon = -\mu_0 \mathbf{H}\cdot \mathbf{M} + ( \mu_0  M_{\rm s}^2 /2  - K_{\rm u}) \sin^2 \theta$, with $\mathbf{H}$ the applied magnetic field, $\mathbf{M}$ the sample magnetization, $\theta$ the angle between the magnetization direction and the (001) plane, and $K_{\rm u}$ the out-of-plane uniaxial anisotropy constant (see Fig.~\ref{fig:Coord}).

\subsubsection{Brillouin Light Scattering Experiments}

Brillouin light scattering (BLS) measurements (Fig.~\ref{fig:spectre_BLS} and Fig.~\ref{fig:BLS}) were performed focusing about 200~mW of monochromatic light with a wavelength $\lambda=532$~nm onto the sample surface, using a camera objective of numerical aperture 2 and focal length 50 mm. The backscattered light was analyzed using a Sandercock type ($3+3$)-pass tandem Fabry-Pérot interferometer. First, spin wave frequency was measured as a function of the in-plane angle $\phi$ between the applied magnetic field and the $[110]$ Fe direction. The intensity of the magnetic field was set to $\mu_0H = 200$~mT to ensure the in-plane saturation of the sample, while the incidence angle of light was normal to the sample plane in order to probe spin waves with zero wave vector. 
Then, the SW dispersion relation was measured in the magnetostatic surface spin-wave (MSSW) configuration applying an in-plane magnetic field $\mu_0H = 200$~mT and sweeping the in-plane wave vector along the perpendicular direction. Due to the conservation of the in-plane momentum in the scattering process, the wave vector magnitude $k$ is linked to the incidence angle of light $\theta$ by the simple relation $k= \frac{4{\pi}}{{\lambda}}{\sin{\theta}}$.   
BLS spectra (Fig.~\ref{fig:spectre_BLS}) were recorded at different incidence angles of light between $0^{\circ}$  and $70^{\circ}$, corresponding to a variation of the in-plane transferred wave vector from 0 to $2.3\times 10^7$~rad/m.


\section{Theoretical Model}

The measured dispersion relation of the spin waves was evaluated with the Landau-Lifshitz equation translated as an eigenvalue problem $i\omega\mathbf{m}(\mathbf{r})=\gamma \mu_0 \Tilde{\mathbf{A}}\mathbf{m}(\mathbf{r})$, where matrix $\Tilde{\mathbf{A}}$ contains all the magnetic interactions in the system. The theory of spin-wave dynamics in magnetization-graded films was developed by Gallardo \textit{et al.} \cite{Gallardo2019}, where the dynamic matrix method \cite{Henry16} was used to account for the variation of the saturation magnetization along the thickness $d$ for different profiles. 
In this work, the dynamic matrix theory is extended to the case of a thick film, where the anisotropies $K_{\rm u}$ and $K_{\rm c}$ are spatially modulated to study the irradiated Fe:N samples. 
The new matrix elements $A_{\eta_{\nu}\eta_{p}}(\eta=y^{\prime},z^{\prime})$, including the anisotropy terms for the $\nu$-th sublayer, are given by (see details in Ref.~\onlinecite{Gallardo2019})
\begin{equation*}
    \begin{aligned}
        A_{y^{\prime}_{\nu}y^{\prime}_{\nu}}&=A_{z^{\prime}_{\nu} z^{\prime}_{\nu}}=0 ,
        \\
        A_{y^{\prime}_{\nu}z^{\prime}_{\nu}}&=-H-\frac{J}{\mu_0 M_{\rm s}d_s} \sum_{i}(\delta_i^{\nu-1}+\delta_i^{\nu+1})
        \\
        &- \frac{2 A_{\rm ex}}{\mu_0 M_{\rm s}} k^2 -\frac{2 M_{\rm s}}{|k|d_s}\sinh\left(\frac{|k|d_s}{2}\right)e^{-|k|\frac{d_s}{2}}
        \\
        &+ \frac{2 K_{\mathrm{u}_\nu}}{\mu_0 M_{\rm s}}-\frac{2 K_{\mathrm{c}_{\nu}}}{\mu_0 M_{\rm s}},
        \\
        A_{z^{\prime}_{\nu}y^{\prime}_{\nu}}&=H+\frac{J}{\mu_0M_{\rm s}d_s} \sum_{i}(\delta_i^{\nu-1}+\delta_i^{\nu+1})+\frac{2 K_{\mathrm{c}_{\nu}}}{\mu_0 M_{\rm s}}
        \\
        &+\frac{2 A_{\rm ex}}{\mu_0 M_{\rm s}} k^2 + M_{\rm s} - \frac{2 M_{\rm s}}{|k|d_s}\sinh\left(\frac{|k|d_s}{2}\right)e^{-|k|\frac{d_s}{2}}.
    \end{aligned}
\end{equation*}
For different sublayers $\nu$ and $p$ ($\nu\neq p$), the matrix elements are
\begin{equation*}
    \begin{aligned}
         A_{y^{\prime}_{\nu} y^{\prime}_{p}}&=-A_{y_{\nu} y_{p}}=
         \\
         &-ik\frac{2M_{\rm s}}{|k|d_s}\sinh^2\left(\frac{|k|d_s}{2}\right)e^{-|k||\nu-p|d_s} \text{sgn}(\nu-p),
         \\
        A_{y^{\prime}_{\nu}z^{\prime}_{p}}&=\frac{2M_{\rm s}}{|k|d_s}\sinh^2\left(\frac{|k|d_s}{2}\right)e^{-|k||\nu-p|d_s}
        \\ 
        &+ \frac{J}{\mu_0M_{\rm s}d_s} (\delta_p^{\nu-1}+\delta_p^{\nu+1}),
        \\
        A_{z^{\prime}_{\nu}y^{\prime}_{p}}&=\frac{2M_{\rm s}}{|k|d_s}\sinh^2\left(\frac{|k|d_s}{2}\right)e^{-|k||\nu-p|d_s}
        \\ 
        &- \frac{J}{\mu_0M_{\rm s}d_s} (\delta_p^{\nu-1}+\delta_p^{\nu+1}).
    \end{aligned}
\end{equation*}
\noindent Here, $K_{\rm{u}_\nu}$ and $K_{\rm{c}_\nu}$ are the uniaxial and cubic anisotropy constants of sublayer $\nu$, $\delta_{i}^{j}$ is the Kronecker delta function ($\delta_{i}^{j}=0$ for $i\neq j$ and $\delta_{i}^{j}=1$ for $i=j$), sgn$(\xi)=\xi/|\xi|$ is the sign function and $J=2A_{\rm ex}/d_s$ \cite{Gallardo2019,Sluka19}, with $A_{\rm ex}$ being the exchange constant and $d_s$ the thickness of each sublayer. Note that the total thickness of the film is given by $d=N_sd_s$, where $N_s$ is the total number of sublayers.
Then, the eigenvalue problem can be numerically solved with these matrix elements to obtain the frequencies and spatial profiles of the dynamic magnetization that support the experimental data. 
It is worth noting that the orthogonal coordinate system ($x', y', z'$) is chosen such that $z'=z$ is the normal axis, the equilibrium magnetization aligns with $x'$ that points along [100] (easy axis), and $y'$ lies in the film plane.

\section{Results and Discussion}

In the first step, the saturation magnetization (\(M_{\rm s}\)) of all samples was estimated using VSM measurements. Both pristine Fe and the implanted samples exhibited \(M_{\rm s}\) values closely matching the expected value of bcc Iron \cite{Danan1968}, i.e., $M_{\rm s} = 1713$ kA/m, within an error margin of $\pm5\%$ mainly attributed to the evaluation of the thin film volume. 
This confirms previous studies, where it was demonstrated that iron epitaxial thin films grown on ZnSe/GaAs (001) exhibit the expected magnetic moment for bcc iron \cite{Marangolo2004} and that no significant change in $M_{\rm s}$ is detected in Fe:N thin films \cite{theseGarnier}. 

\begin{figure}
    \centering
    \includegraphics[width=.4\paperwidth]{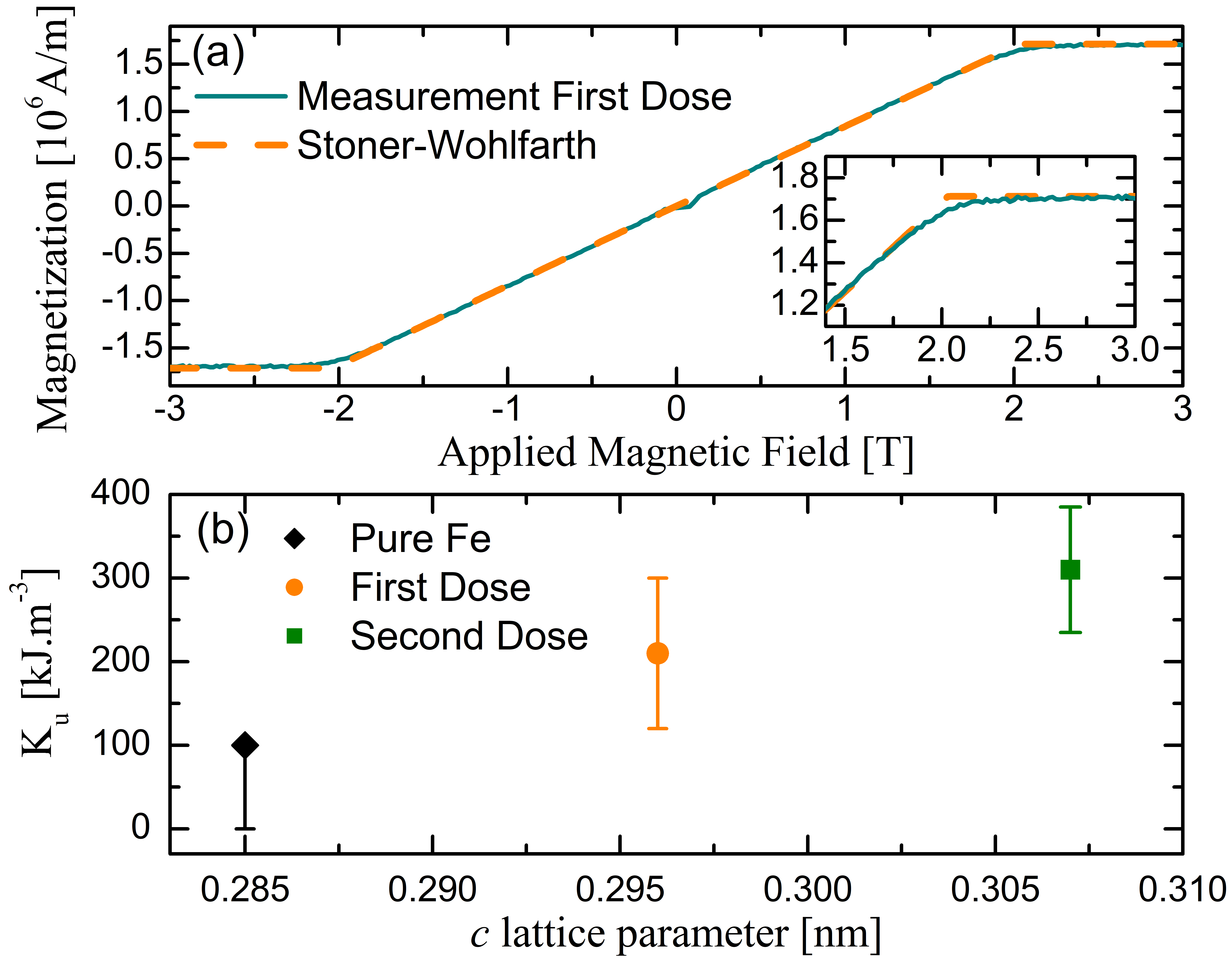}
    \caption{(a) Out-of-plane VSM measurements (First dose: 1.0 $\times 10^{16}$ N\textsubscript{2}/cm\textsuperscript{2}) compared with the Stoner-Wohlfarth model. The inset is a zoom to show the quality of the fit. (b) Out-of-plane anisotropy obtained by fitting the VSM data.     }
    \label{fig:OOPmaille}
\end{figure}

Consequently, the uniaxial anisotropy constant, $K_{\rm u}$, was extracted using the Stoner-Wohlfarth model described above and assuming $M_{\rm s}=1713$ kA/m for all the samples. Figure~\ref{fig:OOPmaille}(a) presents the fit performed for the first dose Fe:N sample, demonstrating the effectiveness of this approach. It was observed that $K_{\rm u}$ increases with increasing irradiation dose. As shown in Fig.~\ref{fig:OOPmaille}(b), which depicts the variation of $K_{\rm u}$ as a function of the measured c-axis lattice parameter, the enhancement of PMA can be attributed to the tetragonal distortion induced by N implantation \cite{Amarouche2017,theseGarnier}. 
The error bars are calculated by considering the lowest and highest values of $M_{\rm s}$ compatible with VSM measurements, i.e., with 
$M_ {\rm s}\in[1627,1799]$ kA/m for Fe:N, and 
$M_ {\rm s}\in[1627,1713]$ kA/m for pristine Fe.

\begin{figure}
    \centering
\includegraphics[width=0.35\paperwidth]{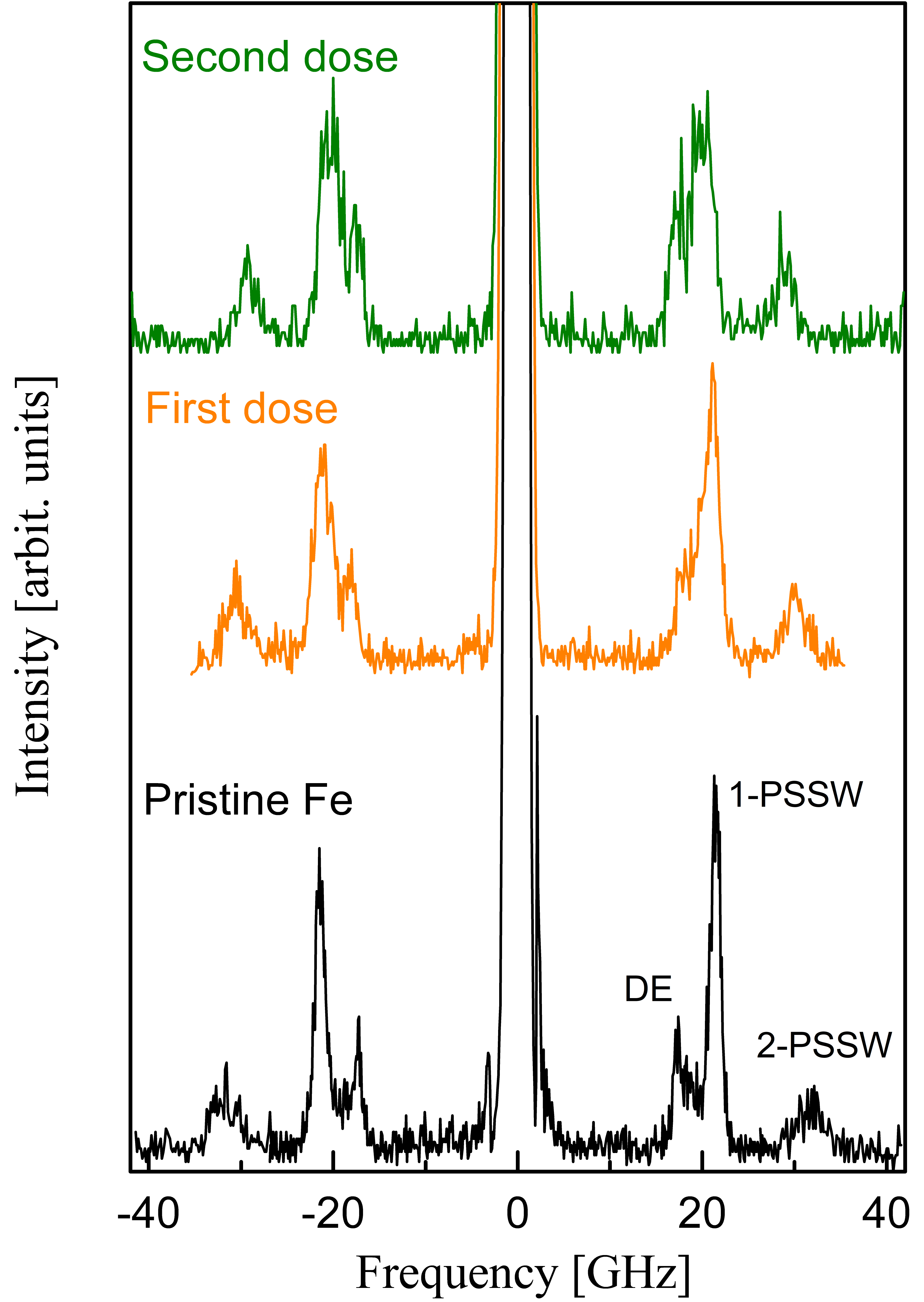}
    \caption{Brillouin light scattering spectra of three investigated samples taken at $k = 0$, applying a magnetic field $\mu_0H = 200$~mT along the [110] direction.}
    \label{fig:spectre_BLS}
\end{figure}

\begin{figure}[h]
  \centering 
  \includegraphics[width=0.45\paperwidth]{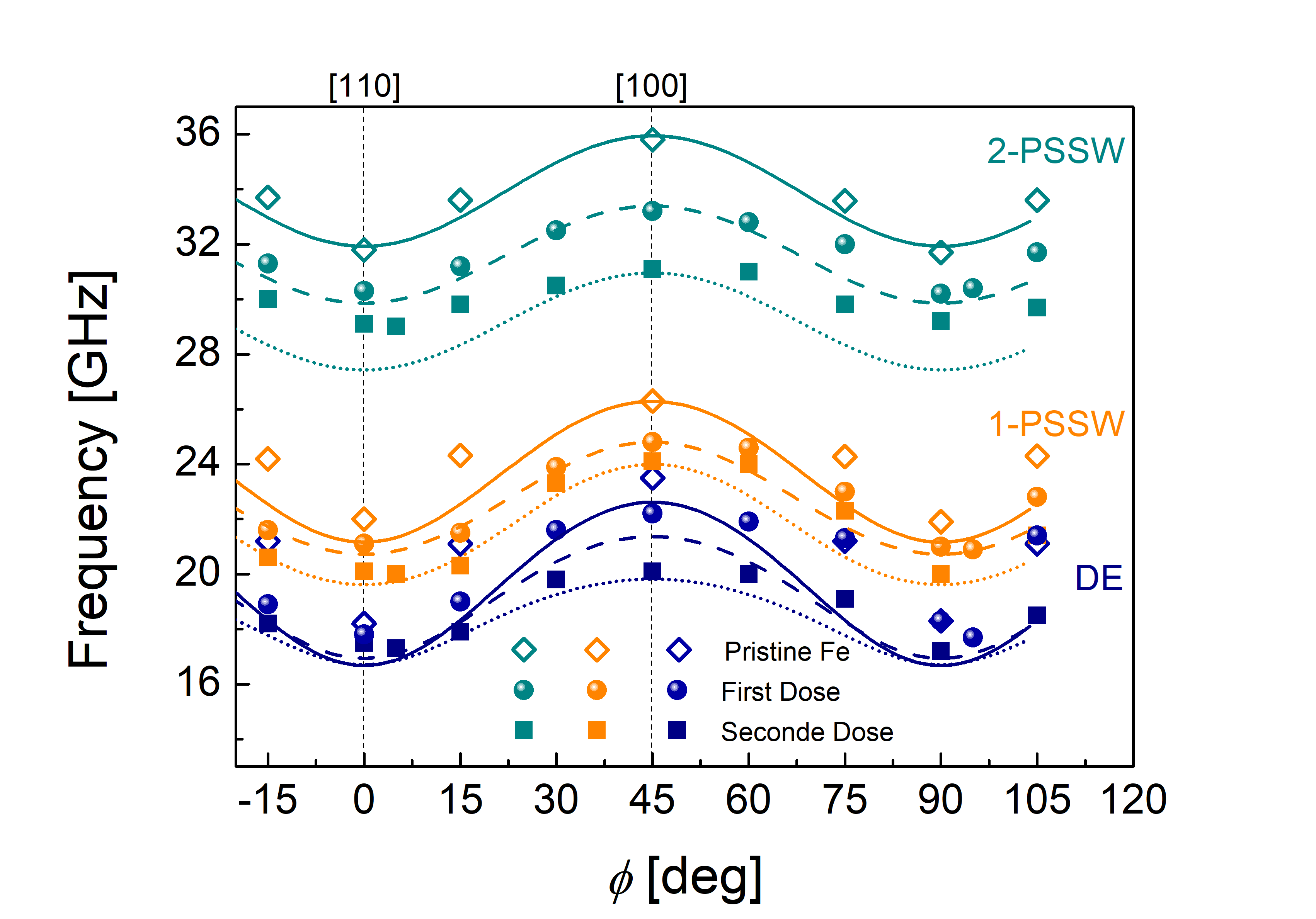}
    \caption{BLS measurements (points) as a function of the in-plane angle $\phi$ between the applied magnetic field and the [110] Fe direction. The solid lines are obtained from the theoretical approach.}
    \label{fig:BLS}
\end{figure}

Then, the dynamic properties of the samples were investigated using BLS measurements. 
Fig.~\ref{fig:spectre_BLS}  reports typical BLS spectra taken for the three investigated samples at $k = 0$  when an in-plane magnetic field $\mu_0 H = 200$~mT is applied along the in-plane [110] direction. Three peaks were observed for both the pristine Fe and the irradiated samples. As discussed in more detail later, the peak at the lowest frequency corresponds to the Damon-Eshbach (DE) mode ($n = 0$), having an almost uniform spatial profile through the film thickness. In contrast, the second and third resonance peaks can be identified as the first and second perpendicular standing spin-wave ($n$-PSSW) modes, characterized by $n = 1$ and $n = 2$ nodes across the film thickness, respectively. 
Figure \ref{fig:BLS} shows the dependence of the spin-wave frequency as a function of the in-plane angle $\phi$, measured by BLS for the pristine Fe film (open diamonds) and the two irradiated samples (dots and squares), applying a magnetic field $\mu_0H = 200$~mT. 
As can be seen, the pristine Fe is characterized by a cubic anisotropy typical of the bulk Fe, with the easy and hard axes along the [100] and [110] directions, respectively. In the irradiated samples, all spin wave modes exhibit a reduction of the in-plane oscillation and a shift toward lower frequencies, which become more pronounced with increasing implantation dose. This suggests an increase in out-of-plane anisotropy and a reduction in the cubic one.

\begin{figure*}[t]
    \centering
    \includegraphics[width=.75\paperwidth]{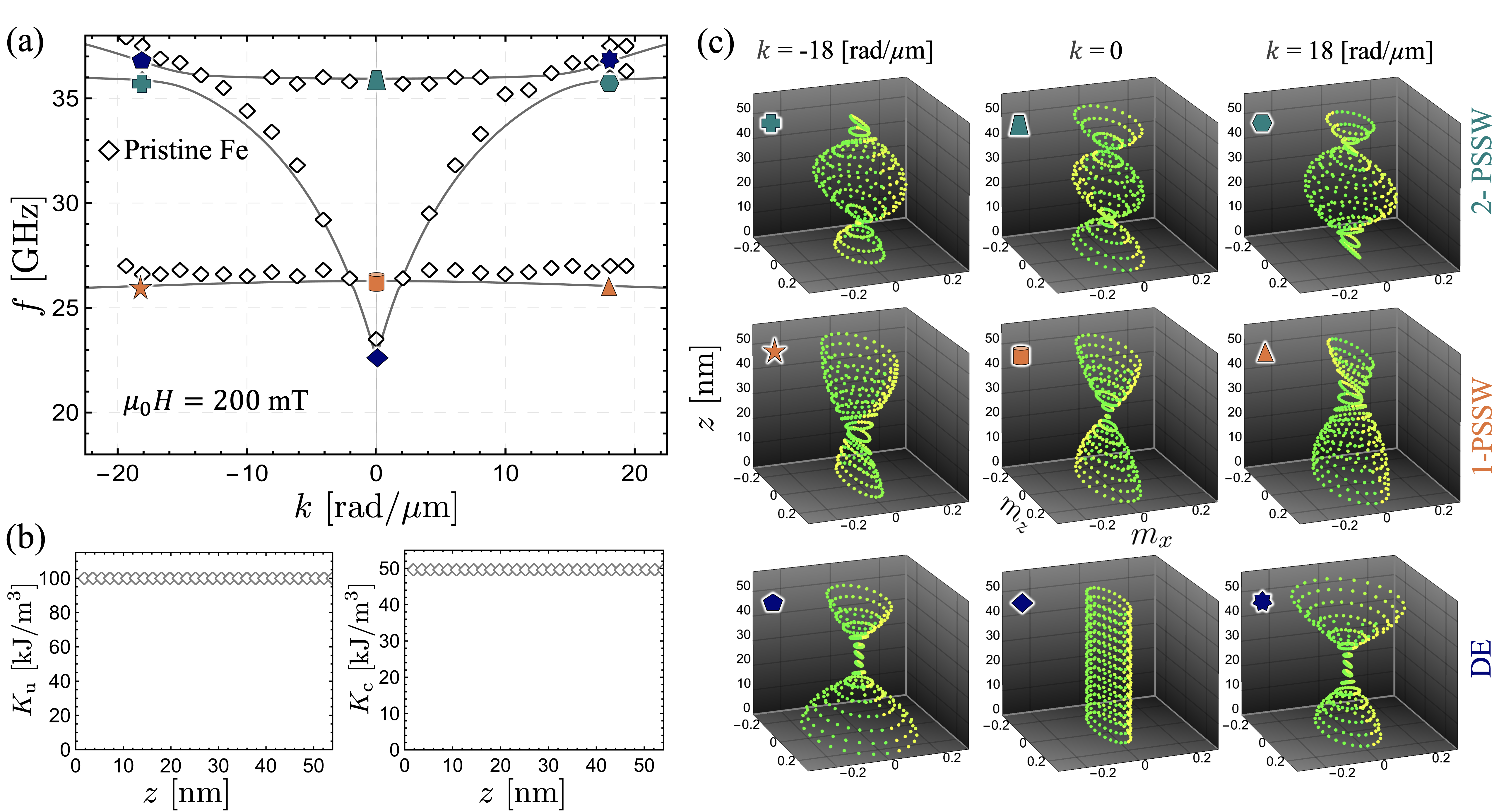}
    \caption{Analysis of the pristine Fe film.  (a) Comparison between the BLS measured (diamonds) and the calculated (continuous lines) frequencies as a function of the wave vector $k$. (b) Values of the magnetic anisotropy constants $K_{\rm u}$ and $K_{\rm c}$  (assuming a constant value along the film thickness for pristine Fe) obtained from the theoretical calculations. Panel (c), precessional amplitudes (in arbitrary units) of the three spin wave modes calculated at  $k=-18$, 0, and $18$ rad$/\mu$m. 
    }
    \label{fig:PureFePr}
\end{figure*}

\begin{figure*}[t]
    \centering
    \includegraphics[width=.75\paperwidth]{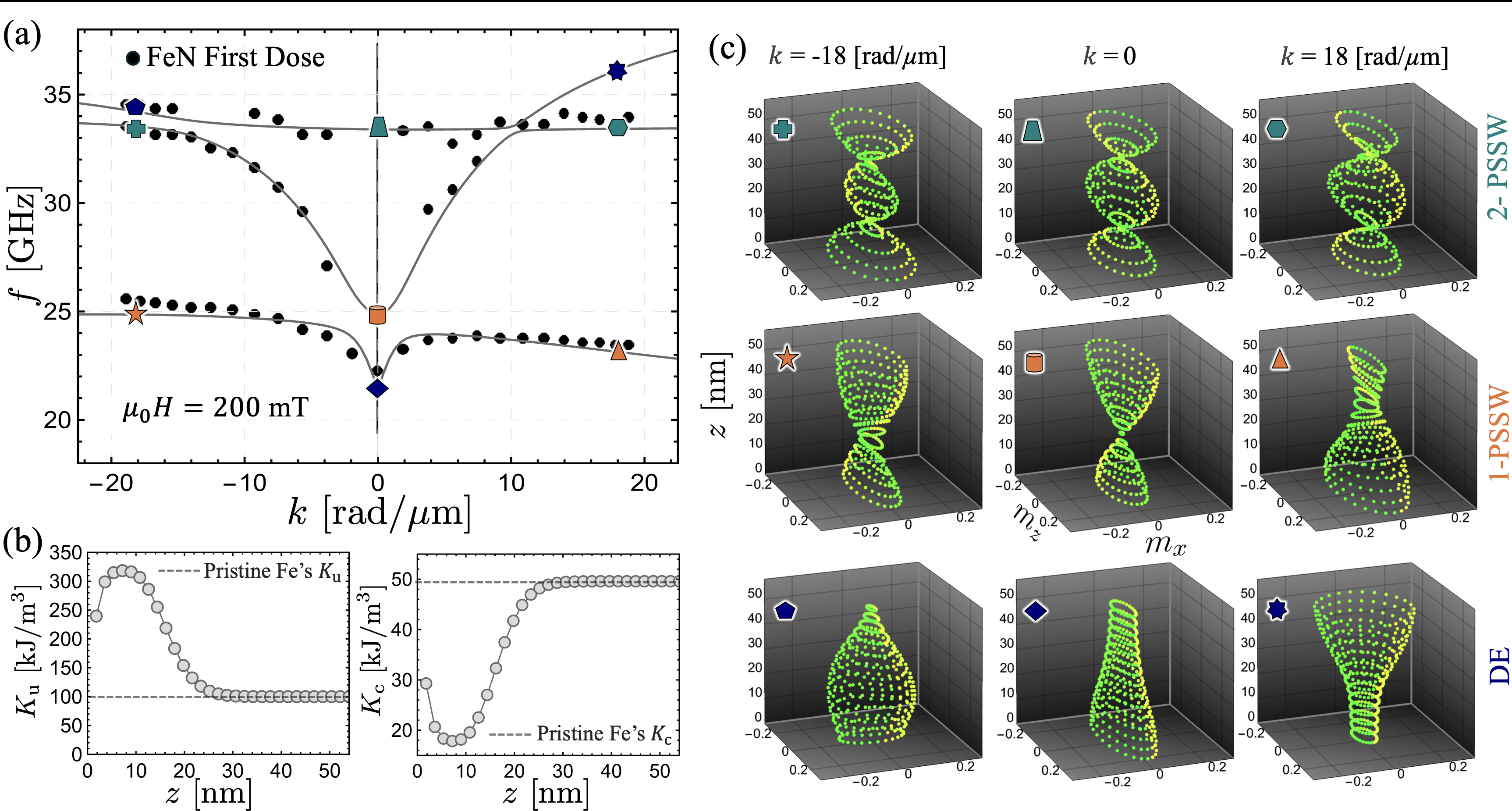}
    \caption{Analysis of the Fe:N film irradiated with the first dose. (a) Comparison between the BLS measured (dots) and the calculated (continuous lines) frequencies as a function of the wave vector $k$. (b) Evolution of the magnetic anisotropy constants $K_{\rm u}$ and $K_{\rm c}$ along the film thickness obtained from the theoretical analysis. Gray dashed lines indicate the value of the anisotropy constant of pristine Fe \cite{Rovillain2022}. (c) Precessional amplitudes (in arbitrary units) of the three spin wave modes calculated at  $k=-18$, 0, and $18$ rad$/\mu$m. }
    \label{fig:FrstDs}
\end{figure*}

\begin{figure*}
    \centering
    \includegraphics[width=.75\paperwidth]{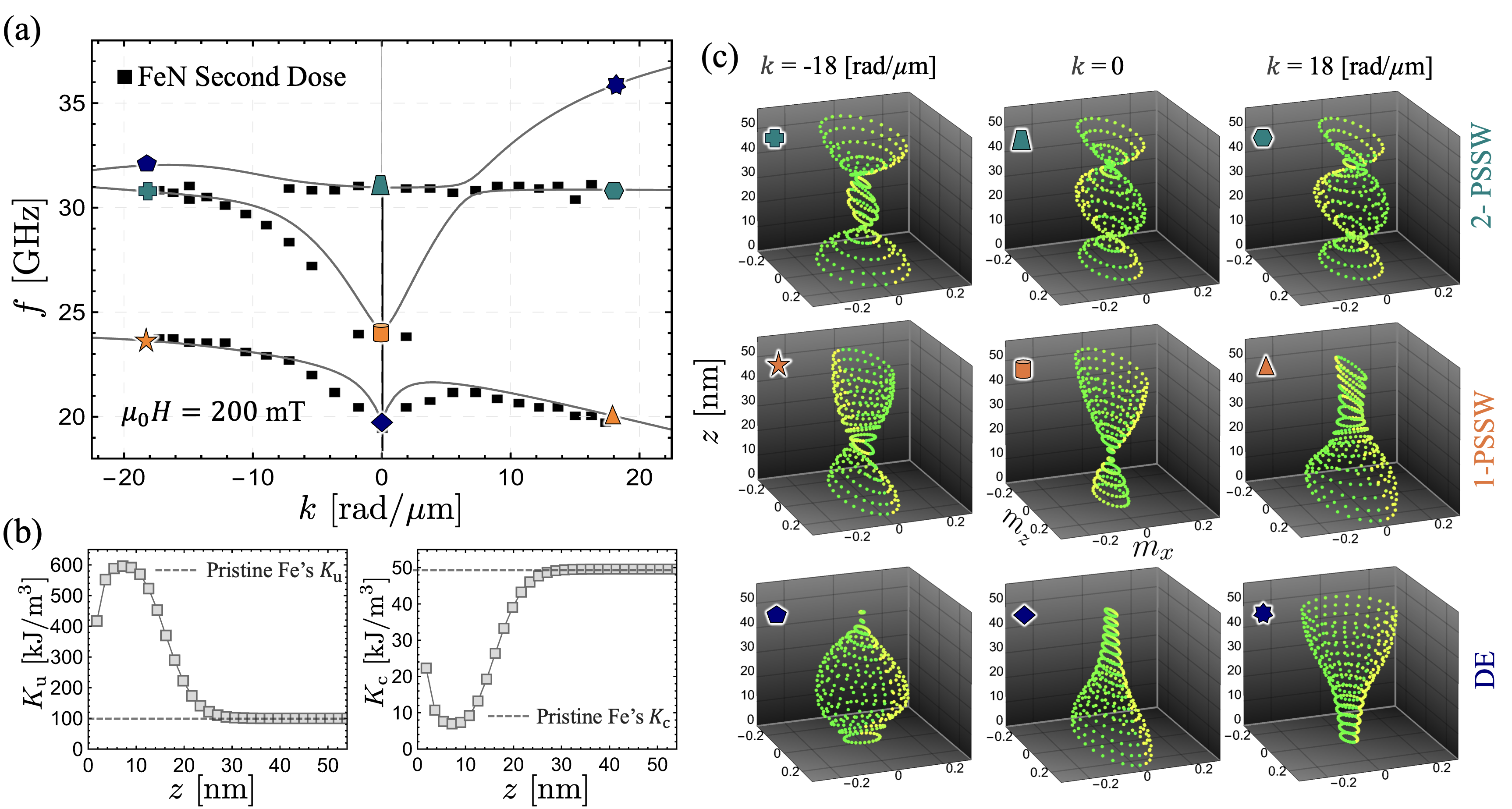}
    \caption{Analysis of the Fe:N film irradiated with the second dose. (a) Comparison between the BLS measured (squares) and the calculated (continuous lines) frequencies as a function of the wave vector $k$. (b) Evolution of the magnetic anisotropy constants $K_{\rm u}$ and $K_{\rm c}$ along the film thickness obtained from the theoretical fit of the data. Gray dashed lines indicate the value of the anisotropy constant of pristine Fe \cite{Rovillain2022}. (c)  Distribution of precessional amplitudes (in arbitrary units) along the thickness of the three spin wave modes calculated at  $k=-18$, 0, and $18$ rad$/\mu$m.}
    \label{fig:ScndDs}
\end{figure*}

Figs.~\ref{fig:PureFePr}-\ref{fig:ScndDs}(a) show the spin-wave dispersion relation measured for the pristine Fe (Fig.~\ref{fig:PureFePr}) and the two irradiated samples (Figs.~\ref{fig:FrstDs} and \ref{fig:ScndDs}). The experiments were performed in MSSW geometry, with a magnetic field applied along the in-plane easy direction [100]. Positive (negative) values of $k$ correspond to the Stokes (anti-Stokes) side of the experimental BLS spectra. 
In order to gain better insight into these findings, the dynamic-matrix approach previously described was exploited to calculate the spin-wave frequency (solid lines) and the spatial profiles through the film thickness of the dynamic magnetization components for each mode [Figs.~\ref{fig:PureFePr}-\ref{fig:ScndDs}(c)]. For the sample of thickness $d=54$ nm, a number of sublayers $N_s=30$ was used, where a convergence of the modes was obtained.

First, the experimental results of the pristine Fe were analyzed (Fig.~\ref{fig:PureFePr}), assuming uniform magnetic parameters through the film thickness. The saturation magnetization was set to $M_{\rm s}=1713$~kA/m. As it can be seen in Figs.~\ref{fig:BLS} and \ref{fig:PureFePr}(a), both the dispersion relation and measurements as a function of $\phi$  are well reproduced for constants values of the anisotropy $K_{\rm c} = 49.6$~kJ/m$^{3}$, $K_{\rm u}=100$~kJ/m$^{3}$ [see Fig.~\ref{fig:PureFePr}(b)], and $A_{\rm ex} = 19$~pJ/m. 
All magnon modes exhibit a reciprocal dispersion relation since anisotropy is uniform across the thickness. One can note that a dispersionless behavior characterizes both the 1-PSSW (orange) and 2-PSSW (green) modes. At the same time, the dispersive curve corresponds to the DE mode (blue), which is coupled with the 1-PSSW and 2-PSSW mode at approximately $2$~rad/$\mu$m and $18$~rad/$\mu$m, respectively.
Fig.~\ref{fig:PureFePr}(c) reports the spatial profiles of the dynamical magnetization calculated for selected values of the in-plane wave vector $k$, corresponding to relevant points of the calculated spin-wave dispersion. 
For $k = 0$, the lowest frequency mode (diamond) corresponds to the uniform one. In contrast, the second and third modes are characterized by one node (cylinder) and two nodes (trapeze) across the thickness, respectively, as expected for the 1-PSSW and 2-PSSW modes. As the wave number $k$ increases, the DE mode assumes a surface character. It localizes at the bottom ($z=54$~nm) and top ($z=0$) surfaces for positive and negative wave vectors, respectively, reflecting the well-known amplitude nonreciprocity of the DE mode \cite{Damon61}. In addition, one can observe that its profile is hybridized with that of the 1-PSSW and the 2-PSSW modes. The latter, indeed, maintain their oscillating behavior, but have a different amplitude at the two film surfaces.

The experimental results of the two irradiated samples were then analyzed (Figs.~\ref{fig:FrstDs} and \ref{fig:ScndDs}). Based on the measurements performed as a function of the in-plane direction of the applied magnetic field, indicating that both the in-plane and the out-of-plane anisotropies are affected by the N implantation, a graded profile that mimics the TRIM simulations (Fig.~\ref{fig:TRIM}) was assumed for both $K_{\rm u}$ and $K_{\rm c}$. 
In particular, $K_{\rm u}$  ($K_{\rm c}$) was set to increase (decrease) in the irradiated part of the film ($z=0$), with the maximum (minimum) value being reached at the depth where a maximum number of implanted N ions is expected, corresponding to 7.2 nm and 7.5 nm for the first and second doses, respectively. Also, for both implantation doses, $K_{\rm u}$ and $K_{\rm c}$ were assumed to recover the values of the pristine Fe at approximately a depth of 25 nm from the surface plane, as can be seen in the graded-anisotropy profiles shown in Figs.~\ref{fig:FrstDs}(b) and \ref{fig:ScndDs}(b). 
Saturation magnetization was assumed to be uniform through the thickness and set to the value $M_{\rm s}=1713$ kA/m. 
Besides, a slight decrease in the exchange constant $A_{\rm ex}$ was required to improve the fitting of the experimental data. 
Table \ref{tab:parametrem} summarizes the parameters used to fit the data.
In contrast to the pristine Fe [Fig.~\ref{fig:PureFePr}(a)], the dispersion relation for the irradiated samples  [Figs.~\ref{fig:FrstDs}(a) and \ref{fig:ScndDs}(a)] shows a noticeable nonreciprocal behavior, which becomes more marked when the implantation dose increase. This nonreciprocal effect makes that the 1-PSSW mode assumes a dispersive behavior with a negative (positive) dispersion for positive (negative) wave vectors.  
In addition, the crossing between the DE and the 1-PSSW mode (2-PSSW) shifts towards smaller wave vector values for $k>0$.

\begin{table*}
\fontsize{10pt}{10pt}
\caption{Magnetic parameters. The uniaxial anisotropy values correspond to the peaks of the graded anisotropy profiles of the irradiated samples. For the cubic anisotropy, the minimum values of the profiles are reported. }
\renewcommand{\arraystretch}{1.5} 
\centering
\begin{ruledtabular}
\begin{tabular}{rccccl}
&$M_{\rm s}$ (kA/m)& $A_{\rm ex}$ (pJ/m)&$K_{\rm c}$ (kJ/m\textsuperscript{3})&$K_{\rm u}$ (kJ/m\textsuperscript{3})  &$d$ (nm) \\
\hline
  Pristine Iron&1713&20&49.6&100  &54\\
        First Dose&1713&17&17.8&318  &54\\
        Second Dose&1713&14&6.9&596  &54\\\end{tabular}
\end{ruledtabular}
\label{tab:parametrem}
\end{table*}

As it can be seen in Fig.~\ref{fig:BLS} and in the dispersion relations in Figs.~\ref{fig:FrstDs}(a) and \ref{fig:ScndDs}(a), a good agreement between BLS data and the theoretical calculation was obtained for both samples. The spatial graduation profiles of $K_{\rm u}(z)$ and $K_{\rm c}(z)$ through the film thickness are illustrated in the panels (b) of Figs.~\ref{fig:FrstDs} and \ref{fig:ScndDs}. A rapid evolution of both the perpendicular and the in-plane anisotropy with the implantation dose can be observed. 
In particular, the perpendicular anisotropy constant is found to increase up to a maximum value of 318~kJ/m$^3$ and 596~kJ/m$^3$ for the first and second dose, respectively. Conversely, the cubic anisotropy constant decreases, reaching the minimum value of 17.8~kJ/m$^3$ for the first dose and 6.9~kJ/m$^3$ for the second one. Note that the out-of-plane anisotropy averaged over the whole film thickness assumes the values  $\langle K_{\rm u}\rangle= 162$ kJ/m$^3$ and $\langle K_{\rm u}\rangle= 240$ kJ/m$^3$ for the first and second implantation doses, respectively, in reasonably good agreement with the values estimated in the VSM analysis. 
As previously discussed, this increase of the PMA with the irradiation dose can be ascribed to the tetragonal distortion of the unit cell caused by the N implantation.

A deeper insight into the influence of the N implantation on the spin-wave dispersion relation can be achieved looking at the panels (c) of Figs.~\ref{fig:FrstDs} and \ref{fig:ScndDs}, which shown the spatial profiles of the dynamical magnetization, calculated for selected values of the in-plane wave vector $k$ according to Figs.~\ref{fig:FrstDs}(a) and \ref{fig:ScndDs}(a).
One can observe that the local variation of the effective magnetization $\mu_0 M_{\rm eff }= \mu_0 M_{\rm s}- 2 K_{\rm u}/M_{\rm s}$ along the film thickness, induced by the graded magnetic anisotropy, strongly affects the spatial profiles of the DE and the 1-PSSW mode.
In particular, at $k=0$, the lowest frequency mode corresponds to the mode without nodes along the thickness as for the pristine Fe. However, it has a nonuniform spatial profile that assumes a larger amplitude in the top region of the film ($z=0$), where $\mu_0 M_{\rm eff }$ has a lower value. 
Also, the spatial profile of the 1-PSSW evaluated at $k=0$ is no longer symmetric through the film thickness and tends to localize in the bottom part of the film ($z=54$ nm) due to the repulsion with the DE mode.
On increasing $k$, the DE mode assumes a surface character. In particular, for positive $k$, it localizes at the bottom region of the film, where $\mu_0 M_{\rm eff}$ has a higher value. In contrast, for negative $k$, it localizes in the top region where $\mu_0 M_{\rm eff }$ assumes a lower value. This behavior explains the marked nonreciprocal dispersion of this mode, which assumes higher (lower) frequency values for positive (negative) wave vectors. It can also be observed that the profile of the DE mode hybridizes with that of the 1-PSSW mode at small wave vector values. The latter, indeed, tends to localize on the opposite surface with respect to the DE mode, resulting in an opposite nonreciprocity, where the 1-PSSW modes at $k<0$ oscillate at higher frequencies compared to the modes evaluated at $k>0$. 
In contrast, the spatial profile of the 2-PSSW mode is only slightly affected by the N implantation. Consequently, this mode exhibits a dispersionless behavior similar to that of the pristine Fe.

Finally, it is interesting to note that the more pronounced nonreciprocity observed for the second dose can be explained considering that this sample is characterized by a more significant local variation of $\mu_0 M_{\rm eff }$ through the thickness due to the larger change of the perpendicular anisotropy induced by the N implantation. 
This result emphasizes the potential for tailoring spin-wave properties through controlled modifications of magnetic anisotropy, which is essential for the design of tunable magnonic devices.

\section{Conclusions}
\label{sec:Conclusions}

In this work, spin wave dynamics in the nitrogen-implanted Fe films have been investigated by combining BLS measurements and theoretical calculations. We show that the significant tetragonal distortion, induced by the low-dose nitrogen implantation, causes in the surface region of the FeN film a gradual increase of the perpendicular anisotropy and a decrease of the in-plane cubic one. This local variation of the magnetic anisotropies along the film thickness, which becomes more marked on increasing the implantation dose, breaks the symmetry and induces a preferential localization of the spin-wave modes in certain regions of the sample, strongly affecting the magnon coupling and the dispersion relation. In particular, the mode at lower frequencies, having a dispersionless behavior in the pristine Fe, is found to assume a dispersive character with opposite group velocity for counterpropagating spin waves. Furthermore, we find that by modifying the implantation conditions, the anisotropy profile changes, which in turn allows for tuning the dispersion relation of the spin-wave modes.
These results demonstrate that nitrogen implantation is a viable technique for engineering magnetic anisotropy gradients in iron films while preserving the film epitaxy, and manipulating spin-wave propagation properties, which are crucial for magnonic applications.

\section{ACKNOWLEDGMENTS}
J. J. B. acknowledges the support of USM-PIIC Nº 054/2023. Support from FONDECYT, Grants No. 1210607 and 1241589, and Basal Program for Centers of Excellence, Grant AFB220001 (CEDENNA), is highly acknowledged. P. R., S. T. and M. M. acknowledge support from the European Union within the HORIZON-CL4-2021-DIGITAL-EMERGING-01. 
P. R.  and M. M. acknowledge the French National Research Agency, ANR-22-CE24-0015 SACOUMAD.
The authors also acknowledge Paola Atkinson for preparing the GaAs substrate by MBE, Jean-Eudes Duvauchelle and the staff of the MPBT (physical properties – low temperature)
platform of Sorbonne Université for their support.
M.M. and S.T. acknowledge PRIN 2022 “Metrology for spintronics: A machine learning approach for the reliable determination of the Dzyaloshinskii-Moriya interaction (MetroSpin)”, 2022SAYARY.
This work has been funded by the European Union - NextGenerationEU, Mission 4, Component 2, under the Italian Ministry of University and Research (MUR) National Innovation Ecosystem grant ECS00000041 - VITALITY - CUP Nos. J97G22000170005 and B43C22000470005.

\begin{figure}[t]
    \centering \includegraphics[width=.3\paperwidth]{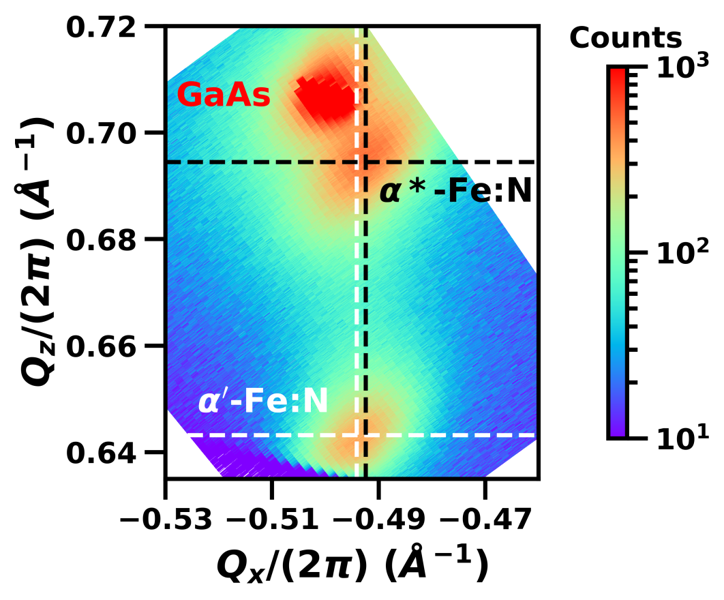}
    \caption{Reciprocal lattice maps showing the (112) diffraction spots of GaAs (or ZnSe) for an implanted sample Fe film ($2\times 10^{16}$ N$_2$ atoms/cm$^2$). The GaAs (or ZnSe) and (112) of $\alpha$’-Fe$_8$N$_{1-x}$ and $\alpha^\star$-FeN$_x$ are clearly observed.}
    \label{fig:DRXmap}
\end{figure}

\begin{figure}[t]
    \centering
\includegraphics[width=.36\paperwidth]{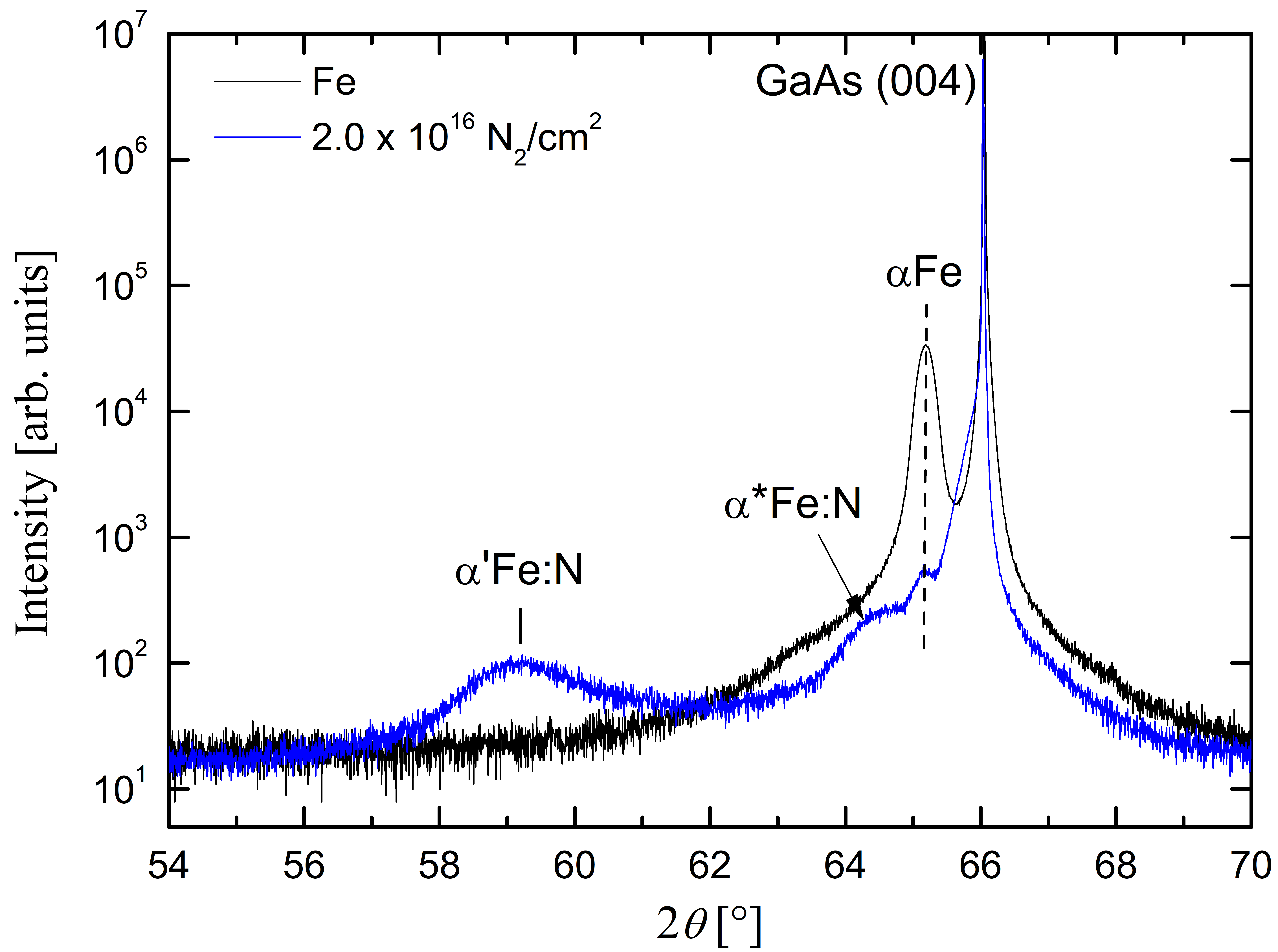}
    \caption{$\theta$-2$\theta$ measurements for the 2.0 $\times$ 10$^{16}$ N$_2$/cm$^2$ Fe:N sample and comparison with pristine Fe. }
    \label{fig:XRD2}
\end{figure}

\appendix
\section{Reciprocal space mapping}
\label{App.A}

A reciprocal space mapping of the 2.0 $\times$ 10$^{16}$ N$_2$/cm$^2$ dose was performed by scanning the $Q_x$ direction (parallel to [110]) and the $Q_z$ direction (parallel to [001]) around the (112) Fe peak. Results are reported in Fig.~\ref{fig:DRXmap}. It is observed that the epitaxial conditions of pristine Fe/GaAs are preserved after N$_2$ implantation. Along the $Q_x$ direction, the (112) reflections of $\alpha'$-Fe$_8$N$_{1-x}$ and $\alpha^*$-FeN$_x$ are located very close to that of $\alpha$-iron within the thin film plane. Along the $Q_z$ direction, while $\alpha^*$-FeN$_x$ is very close to the pristine cubic structure of $\alpha$-Fe, the $\alpha'$ phase exhibits a clear tetragonal distortion. This indicates that nitrogen ion implantation is capable of forming the 
$\alpha'$-Fe$_8$N$_{1-x}$ and $\alpha^*$-FeN$_x$ phases without causing significant modifications to the crystalline structure of the target $\alpha$-Fe, despite the fact that numerous atomic displacements were likely occurring during collision cascades. Thus, the tetragonality of the $\alpha'$-Fe$_8$N$_{1-x}$ phase, discussed in Refs. \onlinecite{Garnier2016, Amarouche2017,theseGarnier}, is confirmed. The in-plane and out-of-plane lattice parameters for all doses are reported in Table \ref{tab:parametrec}. The tetragonality of the $\alpha'$-Fe$_8$N$_{1-x}$ phase reaches values of 7$\%$ and 8.4$\%$, respectively, for second (1.5$\times$10$^{16}$~N$_2$/cm$^2$) and the highest (2.0$\times$10$^{16}$~N$_2$/cm$^2$) implantation doses.
Figure \ref{fig:XRD2} shows the $\theta$-$2\theta$ measurements of the (001) plane for the $2\times 10^{16}$ N$_2$/cm$^2$ sample.

%

%

\end{document}